\documentstyle[prl,aps,psfig]{revtex}

\begin{document}
\draft

\twocolumn[\hsize\textwidth\columnwidth\hsize\csname
@twocolumnfalse\endcsname

\widetext
\title{Long range N\'eel order in the triangular Heisenberg model}
\author {Luca Capriotti${^1}$, Adolfo E. Trumper$^{1,2}$ and Sandro Sorella${^1}$}

\address{ 
${^1}$ Istituto Nazionale di Fisica della Materia and International School for 
Advanced Studies, Via Beirut 4, I-34013 Trieste, Italy \\ 
${^2}$ The Abdus Salam International Centre for Theoretical Physics,
P.O. Box 586, I-34014 Trieste, Italy
} 

\date{\today}
\maketitle

\begin{abstract}
We have studied the Heisenberg model on the triangular lattice
using  several Quantum Monte Carlo (QMC) techniques (up to 144 sites), and 
exact diagonalization (ED) (up to 36 sites).  
By studying the spin gap  as a function of the  system size we have obtained 
a robust evidence for a gapless spectrum, confirming the existence of  
long range  N\'eel order.
Our best estimate is  that in the thermodynamic limit the order parameter 
$m^{\dagger}= 0.41 \pm 0.02 $ is reduced by about 59\% from 
its classical value and the ground state energy per site is   
$e_0=-0.5458 \pm 0.0001$ in unit of the exchange coupling. 
We have identified the important ground state correlations 
at short distance.

\end{abstract}
\pacs{71.10.Fd,71.10.Hf,75.10.Lp}
]

\narrowtext
Historically the antiferromagnetic spin-1/2 Heisenberg model on the
triangular lattice was the first proposed Hamiltonian  for a microscopic 
realization of a spin liquid ground state (GS) \cite{anderson}:
\begin{equation}
\hat{H}=J\sum_{\langle {i},{j} \rangle }
\hat{{\bf {S}}}_{i} \cdot \hat{{\bf {S}}}_{j}~,
\label{heisenberg}
\end{equation}
where $J$ is the nearest-neighbors antiferromagnetic exchange and 
the sum runs over spin-$1/2$ operators. At the
classical level the minimum energy
configuration is the well known $120^{\circ}$ N\'eel state.
The question whether the  combined effect of frustration and quantum
fluctuations favors disordered gapped resonating valence bonds (RVB)
or long range N\'eel type order is still under debate. In fact,
there has been a considerable effort to elucidate the nature of the GS   
and  the results of numerical 
\cite{laughlin,nishimori,huse,singh,bernu,elstner,runge,warman,lhuillier,bonimba},
and analytical\cite{jolicoeur,miyake,chubukov,manuel,azaria} works are controversial. 
From the numerical point of view, ED, which is limited to 
small lattice sizes, provides a very important feature\cite{bernu}: 
the spectra of the lowest energy levels order with increasing total spin, 
a reminiscence of the Lieb-Mattis theorem\cite{lieb} for
bipartite lattices, and are consistent with the symmetry of the 
classical order parameter \cite{bernu}. However, other 
attempts to perform a finite size scaling  study of the order parameter 
indicate a scenario close to a critical one or no magnetic order at 
all\cite{nishimori,runge}.

The variational Quantum Monte Carlo (VMC) allows to extend the numerical 
calculations to fairly large system sizes, 
at the price to make some approximations, 
which are  determined by the quality of the variational wavefunction (WF). 
Many WF have been proposed in the literature\cite{laughlin,huse,lhuillier} 
and the lowest GS  energy estimation  was obtained with the long 
range ordered type.  
In particular, starting from the classical N\'eel state,
Huse and Elser \cite{huse} 
introduced important two and three spin  correlation factors 
in the WF: 
\begin{equation}
|\psi_{\rm V}\rangle= \sum_x \Omega(x)  
\exp \Big(  {\gamma \over 2} \sum_{i,j}
v(i-j) S^{z}_{i} S^{z}_{j} \Big) |x\rangle~,
\label{wfhuse1}
\end{equation}
where $|x\rangle$ is an Ising spin configuration specified by assigning
the value of $S^{z}_{i}$ for each site and 
\begin{equation}
\Omega(x) = T(x)\exp{\Big[{\it i} \frac{2\pi}{3} 
\Big( \sum_{i \in {\rm B}} S_{i}^z
- \sum_{i \in {\rm C}} S_{i}^z \Big) \Big]}
\label{wfhuse2}
\end{equation}  
represents the three sublattices (say A, B and C) classical N\'eel state 
in the $xy$-plane multiplied by the three spin term  
\begin{equation}
T(x) = \exp{\Big({\it i}\,\beta\sum_{\langle i,j,k\rangle} 
\gamma_{ijk} S_{i}^z S_{j}^z S_{k}^z \Big)}~,
\label{wfhuse3}
\end{equation}

\noindent defined by the coefficients $\gamma_{ijk}=0,\pm 1$, appropriately
chosen to preserve the symmetries of the classical N\'eel state, and  by  an overall
factor $\beta$ as discussed in Ref.~\cite{huse}.
Since the Hamiltonian is real and commutes with the $z$-component of the total
spin, $\hat{S}^{z}_{\rm tot}$,
a better variational WF on a finite size
is obtained by taking the real part of Eq.~(\ref{wfhuse1})
projected onto the $S^{z}_{\rm tot} = 0$ subspace.
  
For the two body Jastrow potential $v(r)$ 
 it is also possible to work out 
an explicit Fourier transform $v_q$, based on the consistency with  
linear spin wave (SW) results and a careful
treatment of the singular modes coming from the $SU(2)$ symmetry
breaking assumption\cite{franjic,sw}. This analysis  gives   
$v_q=1-\sqrt{ 1+2\gamma_ q/1-\gamma_q}$ for $q\ne0$ 
and $0$ otherwise, 
where  $\gamma_q {=} \left[ cos\,(q_x)\,+2\,cos\,(q_x/2)
cos\,(\sqrt3\,q_y/2)\right]/3$
and the $q$-momenta  are the ones allowed in a 
finite size with $N$-sites. For a better control of the finite
size effects we have chosen to work with clusters having all the
spatial symmetries of the infinite system \cite{bernu}.

In the square antiferromagnet (AF) 
the classical part by itself determines exactly 
the phases (signs) of the GS  in the chosen basis, the so called 
Marshall sign.  For the triangular case the exact phases are
unknown and the classical part is not enough to fix them correctly.
Therefore, one has to introduce  the three-body
correlations of Eq.~(\ref{wfhuse3}). Although these do not 
provide the exact answer, they
allow to adjust the signs of the WF in a non trivial
way without changing the underlying classical N\'eel order. To this
respect it is useful to define an average sign of the variational 
WF relative to the normalized exact GS  $|\psi_{0}\rangle$ as
\begin{equation}
\langle s \rangle = \sum_{x} |\psi_{\rm 0}(x)|^2 
{\rm sgn}\big(\psi_{\rm V}(x)\psi_{\rm 0}(x)\big)~,
\end{equation}
with $\psi(x) = \langle x|\psi\rangle$.  

We have compared the variational calculation
with the exact GS  obtained by ED 
on the $N = 36$ cluster. For completeness we have considered
the more general Hamiltonian with exchange easy-plane anisotropy $\alpha$,
ranging from the $XY$ case ($\alpha = 0$) to the standard spin isotropic
case ($\alpha = 1$).
As shown in Tab.~\ref{tablesign}, in the variational approach the most 
important parameter, particularly for $\alpha \to 1$,
is  the one, $\beta$, controlling the triplet correlations.
Though the overlap of our best variational WF with the exact GS 
is rather poor, the average sign $\langle s \rangle$ 
is in general very much improved by the  triplet term. 
Our interpretation is that short range many body
correlations are very important to reproduce the relative phases of the 
GS  on each Ising configuration.  
The optimal parameters for our initial guess $\psi_{\rm V}$ of the 
GS $\psi_{0}$ are expected to be very weakly size-dependent 
but they are very difficult to  determine  accurately for large sizes. 
For $\alpha=1$ and $N=36$, 
where ED is still possible, our best guess for the GS WF - 
with the maximum overlap and average sign - is slightly different from 
the one determined with the optimization of the energy.
Since the forthcoming calculations, which significantly improve the VMC, 
are more sensitive to the accuracy of the WF rather than to the one 
of the GS energy, henceforth we have chosen to work with $\beta=0.23$ 
for all the system sizes.
\begin{figure}
\centerline{\psfig{bbllx=80pt,bblly=250pt,bburx=510pt,bbury=545pt,%
figure=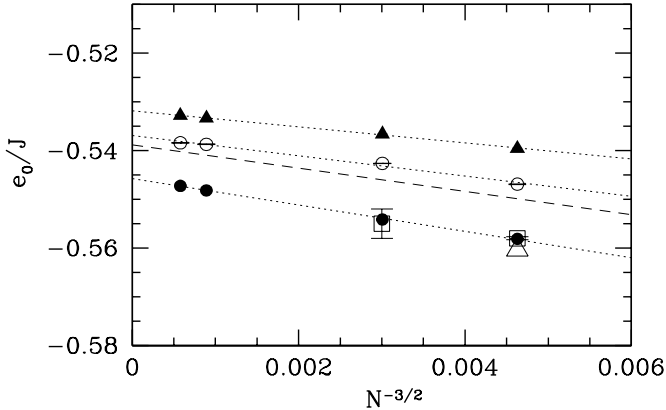,width=80mm,angle=0}}
\caption{\baselineskip .185in \label{sizee}
GS energy per site $e_0 = E_0/N$, in unit of
$J$, as a function of the system size, obtained with
VMC (full triangles), FN (empty dots) and SR with $p = 7$ (full dots) techniques.
SW size scaling \protect\onlinecite{azaria} is assumed and short-dashed lines are
linear fits against $1/N^{3/2}$. The  long-dashed line is the SW prediction,
the empty triangle is the $N = 36$ ED result and the empty squares are data
taken from Ref.~\protect\cite{lhuillier}.
}
\end{figure}

One way to get more accurate GS  properties is to use the
Green Function MC technique (GFMC). As in the fermionic case, for 
frustrated spin systems this numerical method is plagued by the 
well-known sign problem. Recently,
to alleviate the above mentioned instability, 
the Fixed-Node (FN) GFMC scheme \cite{fn} has been introduced  
as a variational technique, typically  much better than the conventional VMC. 
As shown  in Fig.~\ref{sizee}, and also pointed in Ref.~\cite{ss}, for frustrated
spin systems, this technique does not represent a significative advance 
compared to VMC, leading therefore to results biased by the 
variational ansatz.  

\begin{figure}
\centerline{\psfig{bbllx=60pt,bblly=330pt,bburx=540pt,bbury=550pt,%
figure=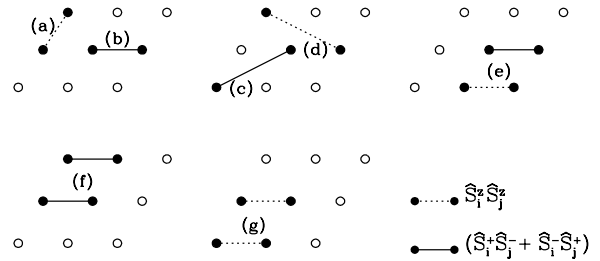,width=80mm,angle=0}}
\caption {\baselineskip .185in \label{corrfun}
Short range spin correlation functions generated by $\hat{H}$ (a,b) and  $\hat{H}^2$ (c-g). 
}
\end{figure}

In order to overcome this difficulty we have   used 
a recently developed technique: GFMC with Stochastic Reconfiguration (SR)
\cite{ss}, which allows to release the FN approximation, 
in a controlled but approximate way, yielding, as shown in Fig.~\ref{sizee}
much lower energies, even for the largest sizes where ED is not possible.
In the appropriate limit \cite{ss} of large number of walkers and high
frequency of SR, the residual bias introduced by the SR depends  only
on the number $p$ of operators used to constrain the GFMC Markov process.  
These  constraints, analogously to the FN one, allow  simulations 
without numerical instabilities.  In principle the exact answer 
can be obtained, within statistical errors, provided $p$ equals 
the huge Hilbert space dimension.
Practically it is necessary to work with small $p$ and an accurate selection 
of physically relevant operators is crucial. As can be easily expected, 
the short range correlation functions $\hat{S}^z_i\hat{S}^z_j$ and
$(\hat{S}^+_i\hat{S}^-_j{+}\hat{S}^-_i\hat{S}^+_j)$ contained 
in the Hamiltonian give  a sizable improvement of the FN GS energy
when they are put  in the SR procedure. 
In order to be systematic we have included in the SR
the short range correlations  generated by $\hat{H}^2$  
(see Fig.~\ref{corrfun}), averaged over all spatial symmetries commuting 
with the Hamiltonian. 
This local  correlations are particularly important to
obtain quite accurate and reliable estimates not only 
of the GS energy but also of the mixed average \cite{ma} of the total spin 
square $\hat{S}^2_{\rm tot}$ and of the order parameter 
$m^{\dagger 2}$ (defined as in Ref.~\cite{bernu}).   
These quantities are easily estimated within the GFMC technique 
and  compared with the exact values computed by ED for  $N=36$
in Tab.~\ref{mixedav}. In particular it is interesting that, starting 
from a variational WF with no definite spin, the GS singlet is
systematically recovered by means of the SR technique. 
Furthermore, as it is shown in Fig.~\ref{sizee},  
the quality of our results is similar 
to the variational one  obtained by P. Sindzingre {\em et al.} \cite{lhuillier},
using a  long range ordered RVB wavefunction.   The latter approach 
is  almost exact for small lattices, but  the sign-problem  is already 
present at the variational level, and the calculation  has not been 
extended to high statistical accuracy or to $N > 48$.

Having obtained an estimate for the GS energy, at least an order of magnitude
more accurate than our best variational guess, it appears possible
to obtain physical features, such as a gap in the spin spectrum,
that are not present  at the variational level.
For instance in the frustrated $J_1{-}J_2$ spin model, with the same technique
and a similar accuracy,  a gap in the spin spectrum
was found in the thermodynamic limit, starting with a similar
ordered and therefore gapless variational WF \cite{ss}.

\begin{figure}
\centerline{\psfig{bbllx=80pt,bblly=250pt,bburx=510pt,bbury=545pt,%
figure=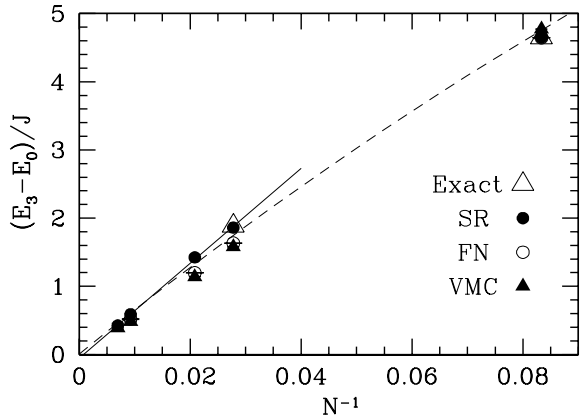,width=80mm,angle=0}}
\caption {\baselineskip .185in \label{gap}
Size scaling of the spin gap to the $S=3$ excitation obtained with VMC, FN and SR (p = 7)
techniques. The long-dashed line is the linear SW prediction and
the solid line is the least-squares fit of the SR data for $N \geq 36$.
}
\end{figure}

In the isotropic triangular AF, the gap to the first
spin excitation is rather small. Furthermore, 
for the particular choice of the guiding WF (\ref{wfhuse1}),
the translational symmetry of the Hamiltonian is preserved only if
projected onto subspaces with total $S^{z}_{\rm tot}$ multiple of three.
Such an $S=3$ excitation belongs to the low-lying states of energy 
$E_S$ and spin $S$ of the ordered quantum AF, behaving as $E_S-E_0 \propto 
S (S+1) /N $\cite{bernu}. 
If instead $E_S-E_0$ remains  finite   for $S=3$ and $N\to \infty$,
this implies a disordered GS.
For all the above reasons we  have studied the gap to the spin $S=3$ excitation 
as a function of the system size.  
As it is shown in Fig.~\ref{gap}, for the lattice sizes for
which a comparison with ED data is possible, the spin gap estimated with the
SR technique is nearly exact. 
The importance to  extend the numerical investigation to clusters 
large enough to allow
a more reliable  extrapolation is particularly evident in the same
 figure in which the $N=12$ and 36 exact data extrapolate linearly to a large 
finite value. This behavior,
is certainly a finite size effect and it is corrected by the SR data 
for $N\ge48$, suggesting, strongly, a  gapless excitation spectrum. 

As we have seen GFMC allows to obtain a very high statistical accuracy on 
the GS energy, but does not allow to compute directly 
GS expectation values 
$\langle\psi_0 | \hat{O} | \psi_0\rangle$\cite{ma}.
A straightforward way is  to perturb the Hamiltonian with 
a term $-\lambda \hat{O}$ , calculate the energy $E(\lambda)$ in presence of 
the perturbation and, by Hellmann-Feynman theorem, estimate 
$ \langle\psi_0 | \hat{O} | \psi_0\rangle = - d E(\lambda) / d \lambda |_{\lambda=0} $ 
with few computations at different {\em small }  $\lambda$'s.
A further complication for non exact calculations like the FN or 
SR, is that if the off-diagonal 
matrix elements $O_{x^\prime,x}$  of the 
operator  $\hat{O}$  (in the chosen basis)   have the opposite sign 
of the product $\psi_{\rm V}(x^\prime) \psi_{\rm V}(x)$, they cannot be handled exactly 
within FN because these matrix elements  change the nodes of $\psi_{\rm V}$.   
A way to circumvent this difficulty if to split the operator 
$\hat{O}$  in three contributions: 
$\hat{O}= \hat{D}+ \hat{O}^{+}+ \hat{O}^{-}$, where $\hat{O}^+$ ($\hat{O}^-$) 
is the operator with the same off-diagonal matrix 
elements  of $\hat{O}$ when they have
the same (opposite) signs of $\psi_{\rm V}(x^\prime) \psi_{\rm V}(x)$, 
and zero otherwise, 
whereas $\hat{D}$ is the diagonal part of $\hat{O}$. 
Then  we can add to the Hamiltonian a contribution that does not change the 
nodes: $\hat{H}(\lambda) =\hat{H} -  \lambda (\hat{D}+ 2\,\hat{O}^+) $ 
for $\lambda > 0$ and 
$\hat{H}(\lambda) =\hat{H}-\lambda (\hat{D} +2\,\hat{O}^-)$ for $\lambda <0$. 
It is easy to show that $\lim\limits_{\lambda \to 0} 
(E(-\lambda)-E(\lambda))/2 \lambda  = \langle\psi_0 | \hat{O} | \psi_0\rangle$.

\begin{figure}
\centerline{\psfig{bbllx=80pt,bblly=195pt,bburx=500pt,bbury=600pt,%
figure=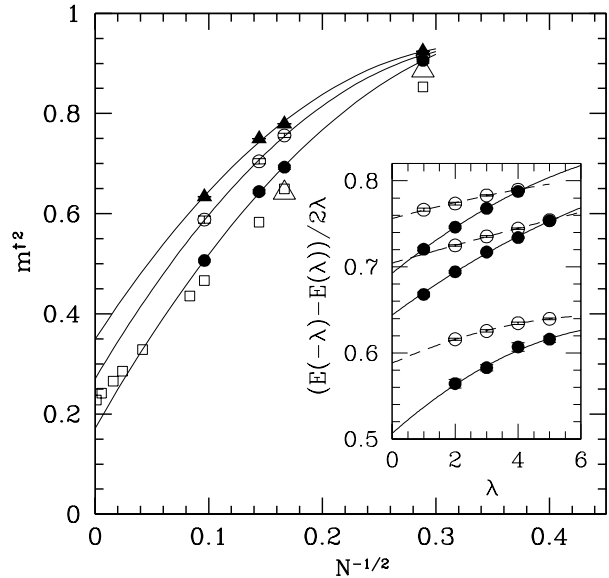,width=80mm,angle=0}}
\caption {\baselineskip .185in \label{mgz}
Size scaling of the order parameter: VMC (full triangles), FN (empty dots),
SR (full dots), exact data (empty triangles) and finite size
linear SW (empty squares). The inset displays the $\lambda \to 0$
extrapolation for $N>12$.  Lines are quadratic fits in all the plots.
}
\end{figure}

We plot in Fig.~\ref{mgz} $m^{\dagger 2}$ 
estimated with this method using the FN and SR techniques. 
For the order parameter the inclusion of many short range correlations in the SR is
not very important (see Tab.~\ref{mixedav}). Then, in order to minimize the numerical effort,
we have chosen to put in the SR conditions the first four correlation functions shown
in Fig.~\ref{corrfun}, the order parameter itself and $\hat{S}^{2}_{\rm tot}$.
While the FN data extrapolate to a value 
not much lower than the variational result, the SR
calculation provides a much more reliable estimate of the order parameter
with no apparent loss of accuracy with increasing sizes.
In this way we obtain for $\hat{m}^{\dagger}$ a value  
well below the linear and the second order (which has actually
a {\em positive} correction \cite{miyake}) SW predictions.
This is partially in agreement with the conclusions of the finite temperature 
calculations ~\cite{elstner} suggesting a GS with a small but nonzero long
range AF order and with series expansions \cite{singh}  
indicating the triangular antiferromagnetic Heisenberg 
model to be likely ordered but close to a critical point. 
However in our simulation, which to our knowledge 
represents a first attempt to perform a systematic finite size scaling 
analysis of the order parameter, the value of $\hat{m}^{\dagger}$ remains 
sizable and finite, consistent with a gapless spectrum. 
This features could be also verified experimentally  
on the K/Si(111):B interface \cite{ksi111} which has turned out
recently to be the first realization of a really bidimensional triangular AF.

Though there is classical long range order, both the VMC and the SR approach
show the crucial role of GS correlations defined on the smallest four spin clusters:
in the variational calculation they are important to determine the correct 
relative phases of the GS WF whereas in the latter more accurate approach this 
correlations allow to obtain very accurate results for the energy and the spin gap 
and to restore the spin rotational invariance of the finite size GS. 

Useful communications with M. Boninsegni and   P. W. Leung 
are acknowledged. We also thank M. Calandra 
and A. Parola for help and fruitful discussions.    
This work was supported in part by INFM (PRA HTSC and PRA LOTUS), 
CINECA grant and CONICET (A.E.T.).

\squeezetable

\begin{table}
\begin{tabular}{dcccccc}
$\alpha$ &  $\beta$   & $\langle s \rangle$ &
$ \langle \psi_0 | \psi_{\rm V} \rangle^{2}$  & $E_{0}/J$ &$\%$ \\
\tableline
0.00&         0.0   &  0.9942 &  0.8610  &  -14.5406  &  1.7 & \\
    &         0.09  &  0.9952 &  0.9303  &  -14.6813  &  0.8 & \\
\tableline
0.50&         0.0   &  0.9100 &  0.5274  &  -16.4229 &   4.0 & \\
    &         0.14  &  0.9597 &  0.6650  &  -16.7016 &   2.4 & \\
\tableline
0.75&         0.0   &  0.8200 &  0.3712  &  -17.5459 &   5.5 & \\
    &         0.17  &  0.9183 &  0.5353  &  -17.9630 &   3.2 & \\
\tableline
1.00&         0.0   &  0.7331 & 0.3157   & -18.5275  &   8.2 & \\
    &         0.19  &  0.9323 & 0.5743   & -19.4400  &   3.6 & \\
    &         0.23  &  0.9372 & 0.6070   & -19.4239  &   3.7 & \\
\end{tabular}
\caption{Average sign, overlap, GS energy and its percentage error 
obtained with the variational WF of Eq.~(\ref{wfhuse1}) for $N = 36$
and some values of the easy-plane anisotropy $\alpha$. 
The  calculations were performed by summing exactly over all 
the configurations. 
}

\label{tablesign}
\end{table}

\begin{table}
\begin{tabular}{dccccccc}
                            &   VMC    &   FN       &  SR($p=2$)  &  SR($p=4$)  &  SR($p=7$) & Exact \\
\tableline
$e_{0}/J$                   & -0.5396  & -0.5469(1) & -0.5534(1)  & -0.5546(1)  & -0.5581(1) & -0.5604\\
$S_{\rm tot}^2$             &  1.71    &  1.20(1)   &  0.65(1)    &   0.46(1)   &  0.02(1)   &   0.00 \\
$m^{\dagger 2}$             &  0.7791  &  0.7701(4) &  0.7659(2)  &  0.7546(3)  & 0.7512(3)  & 0.7394 \\
\end{tabular}
\caption{Variational estimate (VMC) and mixed averages (FN, SR and Exact) 
of the GS energy  per site, of the total spin square and of the AF  
order parameter. SR data are obtained using the first two ($p=2$), 
four ($p=4$) and all ($p=7$) the correlation functions shown 
in Fig.~\ref{corrfun}.
}
\label{mixedav}
\end{table}


\begin{references}
\bibitem{ma} We recall that the mixed average of an operator $\hat{O}$  
is defined as
$ \langle \psi_{\rm V} | \hat{O} | \psi_{0} \rangle / \langle \psi_{\rm V} | \psi_{0} \rangle~,$
where $|\psi_{\rm V}\rangle$ is the guiding 
WF for importance sampling.
\end{references}
\end{document}